\documentclass[
 aip,
 amsmath,amssymb,
 reprint,%
 nofootinbib
]{revtex4-1}

\usepackage{graphicx}
\usepackage{dcolumn}
\usepackage{bm}
\usepackage{enumerate}
\usepackage{siunitx}
\usepackage{amsmath}
\usepackage{cases}
\usepackage{xcolor}

\DeclareSIUnit{\calorie}{cal}
\newcommand{\brho}{\boldsymbol{\rho}}
\newcommand\beq{\begin{equation}}
\newcommand\eeq{\end{equation}}
\renewcommand{\d}{\mathrm{d}}
\newcommand{\q}{\mathbf{q}}
\newcommand{\rr}{\mathbf{r}}
\newcommand{\angstrom}{\text{\normalfont \AA}}

\newcommand{\PR}[1]{{\color{black} #1}}
\newcommand{\LB}[1]{{\color{black} #1}}

\begin{document}


\title{Interaction confinement and electronic screening in two-dimensional nanofluidic channels}

\author{Nikita Kavokine}
\affiliation{Center for Computational Quantum Physics, Flatiron Institute, 162 5$^{ th}$ Avenue, New York, NY 10010, USA}
\author{Paul Robin}%
\author{Lyd\'eric Bocquet}
\email{lyderic.bocquet@ens.fr}
\affiliation{Laboratoire de Physique de l'\'Ecole Normale Sup\'erieure, ENS, Universit\'e PSL, CNRS, Sorbonne Universit\'e, Universit\'e Paris Cit\'e, Paris, France}
\date{\today}

\begin{abstract}
The transport of fluids at the nanoscale is fundamental to \LB{manifold} \PR{biological and industrial} processes, ranging from neurotransmission to ultrafiltration. Yet, it is only recently that well-controlled channels with cross-sections as small as a few molecular diameters became \PR{an experimental reality}. When aqueous electrolytes are confined within such channels, the Coulomb interactions between the dissolved ions are reinforced due to dielectric contrast at the channel walls: we dub this effect `interaction confinement'. Yet, no systematic way of computing these confined interactions has been proposed beyond the limiting cases of perfectly metallic or perfectly insulating channel walls. Here, we introduce a new formalism, based on the so-called surface response functions, that expresses the effective Coulomb interactions within a two-dimensional channel in terms of the wall's electronic structure, described to any desired level of precision. We use it to demonstrate that in few-nanometer-wide channels, the ionic interactions can be tuned by the wall material's screening length. \LB{We illustrate this approach} by implementing these interactions in brownian dynamics simulations \LB{of a strongly confined electrolyte}, \LB{and} show that \LB{the resulting} ionic conduction can be adjusted between Ohm's law and a Wien effect behavior. Our results provide a quantitative approach to tuning nanoscale ion transport through the electronic properties of the channel wall material. 
\end{abstract}

\maketitle

\maketitle

\section{Introduction : the notion of interaction confinement}

Nanofluidic channels -- devices that allow for the well-controlled study of confined water and ion transport -- have experienced a tremendous scale reduction in recent years~\cite{Kavokine2021}. Only a few years ago has molecular scale fluid confinement  been achieved in all possible geometries: 0D nanopores~\cite{Feng2016}, 1D nanotubes~\cite{Tunuguntla2017} and 2D nano-slits~\cite{Radha2016}.  As the size of experimentally accessible channels shrink, the complexity of the theoretical tools required to describe them grows, in particular in what concerns the description of the solid-liquid interface. In microfluidic devices -- with channels on the few micrometer scale -- a wall typically provides a no-slip boundary condition. In nanofluidics -- with channel sizes smaller than 100 nm -- a wall needs to be described in terms of microscopic but still coarse-grained parameters: typically, the surface charge and the hydrodynamic slip length. At the even smaller scale of \emph{single-digit nanopores}~\cite{Faucher2019}, -- channels with one dimension smaller than 10 nm -- we highlight in this paper that the wall needs to be described in terms of its electronic properties, since these affect the inter-particle interactions within the channel. We define \emph{interaction confinement} as the regime where the interactions between particles inside a channel are \PR{modified} by the \PR{presence and }nature of the channel walls. 
 
 Without being named as such, interaction confinement has been known for many years in the theory of biological ion channels. It was realized as early as 1969 by Parsegian~\cite{Parsegian1969} that an ion faces an energy barrier when crossing a \PR{lipid membrane: as the dielectric screening of the ion's Coulomb potential is weaker in the membrane than in water, the ion acquires an additional dielectric self-energy when entering the channel. This corresponds precisely to a modification of the Coulomb potential due to interaction confinement}
 Yet, the equivalence between self-energy barrier and modified inter-particle interactions was realized much later, in the study of strongly confined ion transport~\cite{Cheng2005,Kamenev2006,Zhang2006,Zhang2005,Kaufman2015}. Initial studies focused on one-dimensional tube-like channels; both the channel wall material and the water inside were treated as local dielectric media. The \PR{contrast between the dielectric constants of }the two media was shown to result in the actual confinement of the electric field produced by an ion, with the electric field lines being forced to remain parallel to the channel walls (Fig. 1). The corresponding effective Coulomb interactions are stronger than in bulk water, their strength increasing with decreasing channel width~\cite{Kavokine2021}. In channels with diameters smaller than about 2 nm, these were predicted to cause strong ionic correlations, which in turn result in Bjerrum pairing and ion-exchange phase transitions~\cite{Nicholson2003,Kamenev2006,Zhang2006,Zhang2005}, leading to deviations from Ohm's law transport in the form of Wien effect conduction~\cite{Kavokine2019} and Coulomb-blockade-like phenomena~\cite{Kaufman2015,Kavokine2019}. The concept of reinforced Coulomb interactions due to a dielectric contrast was later extended to a two-dimensional nano-slit geometry~\cite{Robin2021}, where it was found to produce even more striking non-linear ion transport: under the effect of an electric field, the ions may undergo a dynamical phase transition and assemble into dense clusters termed Bjerrum polyelectrolytes~\cite{Robin2021,Zhao2021,Robin2022}.
 
 \begin{figure}
 \centering
 \includegraphics[width=0.45\textwidth]{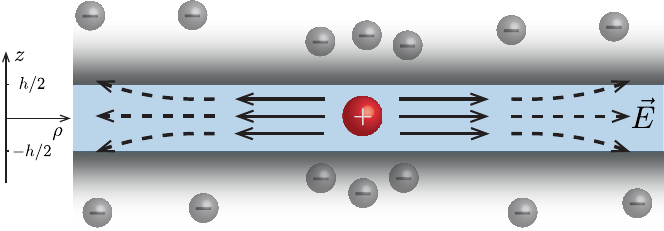}	
 \caption{Schematic view of interaction confinement for a positive ion in a two-dimensional channel. The Coulomb potential produced by the ion is modified by the polarization charges it induces within the channel channel wall. The corresponding field lines are typically confined within the channel instead of pointing isotropically away from the ion.}
 \end{figure}

  The above-mentioned non-linear effects have been studied under the assumption that the channel wall material can be described as a local dielectric medium, with a permittivity much lower than that of bulk water. This is typically the case when the channel is embedded into a lipid membrane, or if it is made of an insulating material such as hexagonal boron nitride (hBN) or $\rm MoS_2$. However, the local dielectric assumption no longer holds for carbon-based materials -- graphene and graphite -- that are widely used to manufacture nanofluidic channels, as these may have conduction electrons 
 \LB{and polarization effects can alter wall-ion interactions \cite{Misra2017,Misra2021}.}
  An opposite approximation, where the channel wall is treated as a perfect metal, has been proposed to study ion \LB{behavior} in carbon nanopores~\cite{Kondrat2011,Lee2014,Merlet2012}. 
  The Thomas-Fermi model may be used to interpolate between a metallic and an insulating behavior~\cite{Mahan} and computations of Coulomb interactions above a Thomas-Fermi surface have been reported~\cite{Vorotyntsev1980,Kornyshev1980,Kornyshev1982,Comtet2017,Kaiser2017,Scalfi2020,Schlaich2022}.
But, to our knowledge, these have not been extended to a confined geometry. \LB{Further,} no practical way of accounting for a more complex dielectric response has been proposed. This is an important shortcoming, since the confined interactions directly control the non-linear ion transport and phase behavior in single-digit nanopores. As a recent example, it was shown experimentally that the confinement-induced shift in the freezing transition of an ionic liquid depends on the metallic or insulating nature of the solid wall \cite{Comtet2017}. 
 
 In this paper, we introduce a method for evaluating the confined Coulomb interactions in a two-dimensional channel with \PR{a slit geometry and} arbitrary wall material, described by its \emph{surface response function}. We derive the general expression \PR{of the confined Coulomb potential} in Sec. II, and in Sec. III we discuss how the surface response function is expressed in terms of a material's electronic properties. In Sec. IV, we evaluate explicitly the confined potential for different channel wall materials. Finally, in Sec. V, we implement confined interactions in Brownian dynamics simulations and show that the ionic conduction in a 2D channel can be adjusted from a Wien effect to an Ohm's law behavior through the electronic properties of the channel wall. 
 
\section{Evaluation of the confined potential}

We consider a point charge $+e$ placed in the middle of a slit-like channel of height $h$ and infinite length and width (Fig. 1). The $z$ axis is perpendicular to the channel walls and we work in cylindrical coordinates $(\brho,z)$. The channel is filled with water, which we for now assume to have a local and isotropic dielectric permittivity $\epsilon_w$. The \PR{dielectric properties} of the channel wall are characterized by the \emph{surface response function} $g(q,\omega)$, which is a well-known quantity in the surface science and plasmonics literature~\cite{Liebsch,Pitarke2007}. It is phenomenologically defined as the reflection coefficient for evanescent plane waves. If an external potential 
\beq 
\phi_{\rm ext}(\brho,z,t) = \phi_0 e^{i(\q \brho -\omega t)}e^{q(z+h/2)} 
\eeq
acts, say, on the confining wall at $z<-h/2$, then the potential induced by the confining wall in the half-space $z>-h/2$ is 
\beq
\phi_{\rm ind}(\brho,z,t) = - \phi_0 g(q,\omega) e^{i(\q \brho -\omega t)}e^{-q(z+h/2)}.
\label{def_g_ph}
\eeq
 In this paper we will only be concerned with static potentials, hence we will use $g(q) \equiv g(q,\omega = 0)$, but general expressions valid for any frequency $\omega$ will still be given where relevant. We discuss in Sec. III how the surface response function is related to the microscopic properties of the wall material. For now, we use it to derive a formal expression for the electrostatic potential within the channel. 
 
 The Coulomb potential inside the \PR{channel} is the sum of the ``external" potential produced by the test charge $+e$, and of the ``induced" potential produced by the polarisation charges in the two confining walls. The external potential is simply the 3D Coulomb potential $\phi_{\rm ext}(\rr) = e/(4\pi\epsilon_0 \epsilon_w | \rr|)$, which supports the following Fourier decomposition: 
\begin{equation}
\phi_{\rm ext}(\rho,z) = \frac{e}{4\pi \epsilon_0\epsilon_w} \int \frac{\d \q}{(2\pi)^2} \frac{2\pi}{q} e^{-q|z|}e^{i \q \brho}.
\label{phiext}
\end{equation}
The induced potential may be determined separately for every wavevector $q$. Let $\phi_{\rm ext}^m(q)$ be the external potential acting on the wall at $z = -h/2$. It is the sum of the potential produced by the test charge, and of the induced potential created by the polarization charges in the medium at $z > h/2$, both screened by the water dielectric constant. By symmetry, the external potential is the same in the upper and lower dielectric medium. This yields the following self-consistent equation: 
\begin{equation}
\phi_{\rm ext}^m (q) = \frac{e}{4\pi \epsilon_0 \epsilon_w} \frac{2\pi}{q} e^{-qh/2} - g(q) \phi_{\rm ext}^m(q) e^{-q h}. 
\label{ion2Dsc}
\end{equation}
We are interested in the total potential in the plane $z = 0$, which is 
\begin{equation}
\phi_{\rm tot} (q,z = 0) = \phi_{\rm ext}(q,z = 0) -2 g(q) \phi_{\rm ext}^m (q) e^{-qh/2}. 
\end{equation}
Making use of eq.~\eqref{ion2Dsc}, we obtain 
\begin{equation}
\phi_{\rm tot}(q,0) =  \frac{e}{4\pi \epsilon_0\epsilon_w} \frac{2\pi}{q} \left( 1- \frac{2 g(q) e^{-qh}}{1+g(q)e^{-qh}} \right). 
\label{main_result}
\end{equation}
The potential in real space is then obtained by inverse Fourier transformation, which thanks to the rotational symmetry reduces to 
\begin{equation}
\phi_{\rm tot} (\rho,0) \equiv \phi(\rho) = \int_0^{+\infty} \frac{\d q}{2\pi} \, q J_0(q \rho) \phi_{\rm tot} (q,0),
\label{invhankel}
\end{equation}
with $J_0$ the Bessel function of the first kind. \LB{These expressions make the link between the confined charge-charge interactions and the surface response fonction of the materials (defined for a semi-infinite medium).}

Eqs.~\eqref{main_result} and ~\eqref{invhankel} constitute the main formal result of this paper.

\section{Surface response functions}

The result in Eqs.~\eqref{main_result} and ~\eqref{invhankel} is of little use unless one is able to evaluate the surface response function for a given channel wall material. This is the purpose of the present section. Throughout this section, we will consider a single interface in the plane $z = 0$, with the solid material filling the half-space $z<0$ and the half-space $z>0$ filled with vacuum. We will subsequently extend our results to the case where the solid is in contact with water.  

\subsection{Case of a local dielectric}

If the solid material is a local dielectric characterized by a permittivity $\epsilon_m$, it cannot contain any induced charges (except on the surface), and the potential therefore solves the Laplace equation $\nabla^2 \phi_m = 0$ inside the material. Upon Fourier transformation, the Laplace equation becomes 
\beq
\PR{\frac{\partial^2 \phi_m}{\partial z^2} (q,z)- q^2 \phi_m(q,z) = 0.}
\label{laplace}
\eeq  
The solution of \eqref{laplace} that vanishes at $z \to -\infty$, is of the form $\phi_m(q,z) = \phi_m e^{qz}$. Outside the material, the potential is the sum of the evanescent wave external potential and the induced potential. Since the Laplace equation holds, the outside potential reads 
\begin{equation}
\phi(q,z) = \phi_{\rm ext} e^{qz} + \phi_{\rm ind} e^{-qz}.
\end{equation}
Now, we must enforce boundary conditions on the interface. These are given by continuity of the potential and of the displacement field $\mathbf{D} = -\epsilon_0 \epsilon \nabla \phi$, where $\epsilon$ is the relative permittivity (1 or $\epsilon_m$). The boundary conditions read
\begin{equation}
\begin{split}
&\phi_{\rm ext} + \phi_{\rm ind} = \phi_m \\
&\phi_{\rm ext} - \phi_{\rm ind} = \epsilon_m \phi_m.
\end{split}
\label{BC_local}
\end{equation}
Hence we obtain $\phi_{\rm ind} = \phi_{\rm ext}(1-\epsilon_m)/(\epsilon_m+1)$, and, using the definition in eq.~\eqref{def_g_ph}, the expression of the surface response function: 
\begin{equation}
g(q) = \frac{\epsilon_m-1}{\epsilon_m +1}. 
\end{equation}
The surface response function thus appears as a generalization of the image charge formalism: in the local dielectric case, its expression corresponds to the magnitude of the image charge~\cite{Jackson}. 

\subsection{General case: microscopic expression}

When no particular assumption can be made for the material's dielectric properties, the surface response function must be determined from its general microscopic expression. For any medium, we may define the density-density response function $\chi$ as the linear response function relating the induced charge density $\delta n$ to the externally applied potential $\phi_{\rm ext}$ (in energy units): 
\begin{equation}
\delta n (\rr,t) = \int_{-\infty}^{+\infty} \d t' \int \d \rr' \chi(\rr,\rr',t-t') \phi_{\rm ext}(\rr',t'). 
\label{chidef}
\end{equation}
In our interface geometry (with translational invariance parallel to the interface), we may define the space-time Fourier transform 
\begin{equation}
\begin{split}
\chi(q,z,z',\omega) &= \int \d(\brho-\brho') \int_{-\infty}^{+\infty} \d(t-t') e^{-i\q (\brho - \brho')} \\ &e^{-i \omega(t-t')}  \chi(\brho-\brho',z,z',t-t').
\end{split}
\end{equation}
The surface response function is then expressed as
\begin{equation}
g(q,\omega) = - \frac{e^2}{2 \epsilon_0q} \int_{-\infty}^{0}\d z \d z' \,  e^{q(z+z')} \chi(q,z,z',\omega). 
\label{gdef}
\end{equation}

We may check that this expression is consistent with the phenomenological definition in eq.~\eqref{def_g_ph}. \PR{Suppose the solid is subject to an evanescent plane wave at frequency $\omega$, of the form $\phi_{\rm ext}(\brho,z,t) = \phi_0 e^{i (\q \brho - \omega t)} e^{q z} $. Its space-time Fourier transform is $\phi_{\rm ext}(z,q,\omega) = \phi_{0} e^{qz}$. Then, the induced charge density is 
 \begin{equation}
\delta n (q,z,\omega) = \phi_{0} \int_{-\infty}^0 \d z' \chi_q(z,z',\omega) e^{qz'},
\end{equation}
and the induced potential at a distance $z$ above the medium is 
\begin{equation}
\begin{split}
\phi_{\rm ind}(q,z,\omega) &= \phi_{0}  \int_{-\infty}^0 \d z' \d z'' \chi(q,z',z'',\omega) \frac{e^2}{2 \epsilon_0 q}  e^{q(z'+z''-z)} \\ &= -g(q,\omega) \phi_{0} e^{-qz},
\end{split}
\end{equation}
or, in real space
\beq
\phi_{\rm ind}(\rho,z,t) = - \phi_0 g(q,\omega_0) e^{i (\q \brho-\omega_0 t)} e^{-q z}.
\eeq
}
Computing the surface response function according to the microscopic expression in eq.~\eqref{gdef} requires the knowledge of the density response function $\chi(q,z,z',\omega)$ for a semi-infinite solid. There exist various analytical and numerical methods for its evaluation to varying degrees of precision. If the solid's polarisation is mainly of electronic origin, the simplest treatment (RPA, Random Phase Approximation) that takes into account electron-electron interactions requires to solve the following integral (Dyson) equation~\cite{Pitarke2007}: 
\begin{equation}
\begin{split}
&\chi(q,z,z',\omega) = \chi^0(q,z,z',\omega) + \dots \\ &+ \int \d z_1 \d z_2 \, \chi^0(q,z,z_1,\omega) V_q(z_1-z_2) \chi (q,z_2,z',\omega), 
\end{split}
\label{dysonchi}
\end{equation}
where $V_q(z) = (e^2/2\epsilon_0 q) e^{-q|z|}$ is the Fourier-transformed Coulomb potential. Here $\chi^0$ is the non-interacting density response function: it determines the electrons' response to an external potential $\phi_{\rm ext}$, the electron-electron interactions being switched off. It can in principle be computed if the eigenenergies $E_{\lambda}$ and eigenfunctions $\psi_\lambda (\rr)$ of the non-interacting system are known~\cite{rammer_ch6}: 
\begin{equation}
\begin{split}
\chi^0(\rr,\rr',\omega) = \sum_{\lambda,\lambda'}\, \frac{n_{\rm F}(E_{\lambda})-n_{\rm F}(E_{\lambda'})}{E_{\lambda}-E_{\lambda'} + \hbar \omega + i \delta} \dots \\ \dots \psi^*_{\lambda}(\rr) \psi_{\lambda'}(\rr) \psi^*_{\lambda'}(\rr') \psi_{\lambda}(\rr'), 
\end{split}
\label{chi0def}
\end{equation}
where $n_{\rm F}$ is the Fermi-Dirac distribution and $\delta \to 0^+$. In this way, the effective Coulomb interactions inside a nanoscale channel are directly related to the channel walls' electronic structure. 

\subsection{Specular reflection approximation}

Even if $\chi_0$ is known, eq.~\eqref{dysonchi} must be solved numerically for every value of $q$ and $\omega$. A considerable simplification is achieved within the so-called specular reflection (SR) approximation~\cite{Griffin1976}, which allows one to solve~\eqref{dysonchi} analytically and express the surface response in terms of the bulk response. The SR approximation sets
\begin{equation}
\chi^0(q,z,z',\omega) = \chi^0_{\rm B} (q,z-z',\omega) + \chi^0_{\rm B} (q,z+z',\omega),
\label{specular}
\end{equation}
where $\chi^0_{\rm B}$ is the bulk system's non-interacting density response. This ansatz does not correspond to any particular form of the wavefunctions in eq.~\eqref{chi0def}. It imposes phenomenologically that in the presence of a surface, the points $z$ and $z'$ may either interact directly, or through a specular reflection from the surface at $z=0$. It can be shown that the SR approximation thus amounts to neglecting quantum interference between electrons impinging on and electrons reflected from the surface~\cite{Griffin1976}. 

Inserting eq.~\eqref{specular} into eq.~\eqref{dysonchi} and carrying out Fourier transforms along the vertical direction (the computation is detailed, for example, in ref.\cite{Griffin1976}), one obtains: 
\begin{equation}
\begin{split}
&g (q,\omega) = \frac{1-q\ell_q(\omega)}{1+ q\ell_q(\omega)}, ~~~ \rm with \\
&\ell_q(\omega) = \frac{2}{\pi} \int_0^{+\infty} \frac{\d q_z}{(q^2+q_z^2)\epsilon(q,q_z,\omega)}
\end{split}
\label{gspecular}
\end{equation}
where $\epsilon(q,q_z,\omega) = 1 - \frac{e^2}{\epsilon_0 (q^2+q_z^2)} \chi^0_{\rm B}(q,q_z,\omega)$ is the bulk system's dielectric function. 

The bulk non-interacting density response function is obtained from the Fourier-transformed eq.~\eqref{chi0def}:
\begin{equation}
\begin{split}
\chi^0_{\rm B} (q,q_z,\omega) =  \sum_{\nu,\nu'} \int_{\rm BZ} \frac{\mathrm{d}^3k}{4\pi^3} |\langle \mathbf{k} + \mathbf{q},\nu | e^{i \mathbf{q \cdot r}}| \mathbf{k} ,\nu'\rangle|^2 \\
\frac{n_{\rm F}[E_{\nu}(\mathbf{k+q})]-n_{\rm F}[E_{\nu'}(\mathbf{k})]}{E_{\nu}(\mathbf{k}+\mathbf{q})-E_{\nu'}(\mathbf{k})-\hbar (\omega +i\delta) },
\end{split}
\label{chi0bulk}
\end{equation}
where we have re-labeled the states $\lambda \mapsto (\mathbf{k},\nu)$, with $\nu$ a band index and $\mathbf{k}$ a vector within the (three-dimensional) first Brillouin zone. 

We report here an alternative derivation of eq.~\eqref{gspecular}, which has the advantage of being computationally simpler than the one reported in \cite{Griffin1976}. It is based on the work of Ritchie and Marusak~\cite{Ritchie1966}, who first proposed the SR approximation in their study of surface plasmons.  The idea is that, when eq.~\eqref{specular} is enforced, the \emph{shape} of the density response of the semi-infinite medium to the potential $\phi_{\rm ext}(q,z,\omega) = \phi_{\rm ext}e^{qz}$ is the same as the \emph{shape} of the density response of an infinite medium to a symmetrized potential $\phi_{\rm eff} (q,z,\omega) = \phi_{\rm eff} e^{-q|z|}$. The amplitude $\phi_{\rm eff}$ is a priori non known, and it is determined by enforcing Maxwell boundary conditions at the interface. 
\begin{widetext}
In the following, we will drop the frequency $\omega$ which plays no role in the computation. In response to the potential $\phi_{\rm eff}$, the induced charge density in the infinite medium reads
\begin{align}
\delta n (q,z) &= \phi_{\rm eff} \int_{-\infty}^{+\infty} \d z' \chi_{\rm B}(q,z-z') e^{-q|z'|}\\
& = \phi_{\rm eff}\frac{1}{2\pi} \int_{-\infty}^{+\infty} \d q_z \chi_{\rm B}(q,q_z) e^{iq_z z} \int_{-\infty}^{+\infty} \d z' e^{-q|z'|} e^{-i q_z z'} \\
& = \phi_{\rm eff} \frac{q}{\pi} \int_{-\infty}^{+\infty} \d q_z \frac{\chi_{\rm B}(q,q_z)}{q^2+q_z^2} e^{i q_z z}.
\end{align}
The induced potential $\phi_{\rm ind,m}$ (not to be confused with the induced potential $\phi_{\rm ind}e^{-qz}$ outside the medium) is 
\begin{align}
\phi_{\rm ind,m} (q,z)& = \int_{-\infty}^{+\infty} \d z' \, \delta n (q,z') \frac{e^2}{4\pi \epsilon_0} \frac{2\pi}{q} e^{-q|z-z'|} \\
& = 2 \phi_{\rm eff} \frac{e^2}{4 \pi \epsilon_0} \int_{-\infty}^{+\infty} \d q_z \frac{\chi_{\rm B}(q,q_z)}{q^2+q_z^2} e^{i q_z z} \int_{-\infty}^{+\infty} \d z' e^{-q |z-z'|} e^{i q_z z'} \\
& = 4 \phi_{\rm eff} \frac{e^2}{4\pi \epsilon_0} \int_{-\infty}^{+\infty} \d q_z \frac{q \chi_{\rm B}(q,q_z)}{(q^2+q_z^2)^2} e^{i q_z z}. 
\label{phiind}
\end{align}
\end{widetext}
At this point, we may introduce the bulk dielectric function $\epsilon(q,q_z)$. For the bulk interacting density response function, the RPA Dyson equation~\eqref{dysonchi} reduces to 
\begin{equation}
\chi_{\rm B}(q,q_z) = \frac{\chi^0_{\rm B}(q,q_z)}{1-\frac{e^2}{\epsilon_0 (q^2+q_z^2)}\chi^0_{\rm B} (q,q_z)}. 
\end{equation}
The dielectric function being defined according to $\epsilon(q,q_z) = 1 - \frac{e^2}{\epsilon_0 (q^2+q_z^2)} \chi^0_{\rm B}(q,q_z)$, we have the relation
\begin{equation}
\chi_{\rm B}(q,q_z) = \frac{\epsilon_0 (q_z^2+q^2)}{e^2} \left( \frac{1}{\epsilon(q,q_z)}-1 \right). 
\end{equation}
When inserting this relation into eq.~\eqref{phiind}, we need to compute the integral 
\begin{equation}
I(q) = \int_{-\infty}^{+\infty} \d q_z\, \frac{e^{i q_z z}}{q^2+q_z^2} = \frac{1}{q} \int_{-\infty}^{+\infty} \d u \, \frac{e^{iu qz}}{1+ u^2}.
\end{equation}
Specializing to the case $z<0$, and noticing that the integrand has poles at $i$ and $-i$, we may close the integration path in the lower complex plane, so that 
\begin{equation}
I(q) = - \frac{2 i \pi}{q} \underset{u= -i}{\mathrm{Res}} \left[\frac{e^{iu qz}}{1+ u^2} \right] = \frac{\pi}{q} e^{qz}. 
\end{equation}
Finally, 
\begin{equation}
\phi_{\rm ind,m} (q,z) = \phi_{\rm eff} \left(\frac{q}{\pi} \int_{-\infty}^{+\infty} \frac{e^{i q_z z}}{(q^2+q_z^2) \epsilon(q,q_z)} -e^{qz} \right), 
\end{equation}
so that the total potential in the half-space $z<0$ is 
\begin{align}
\phi_m(q,z) &= \phi_{\rm eff} e^{qz} + \phi_{\rm ind,m}(q,z) \\
&= \phi_{\rm eff} \, \frac{q}{\pi} \int_{-\infty}^{+\infty} \frac{e^{i q_z z}}{(q^2+q_z^2) \epsilon(q,q_z)}. 
\end{align}
We now need to determine $\phi_{\rm eff}$ in the actual semi-infinite medium by enforcing the boundary conditions at the surface, which are, as in the local dielectric case (section III.A), continuity of the potential and of the displacement field. Outside the medium, we may still express the potential as $\phi_{\rm ext} e^{qz} + \phi_{\rm ind} e^{-qz}$: the sum of the actual potential we are applying and the potential induced by the medium. The displacement field is produced only by the external charges, hence $\mathbf{D}(q,z) = - \epsilon_0 \nabla \phi_{\rm eff} (q,z)$ in the half-space $z<0$, so that the boundary conditions read:
\begin{equation}
\begin{split}
&\phi_{\rm ext} + \phi_{\rm ind} = q \ell_q \phi_{\rm eff} \\
&\phi_{\rm ext} - \phi_{\rm ind} =  \phi_{\rm eff}.
\end{split}
\label{BC_specular}
\end{equation}
We deduce
\begin{equation}
\phi_{\rm ind} = \frac{q \ell_q-1}{q \ell_q +1} \phi_{\rm ext}, 
\end{equation}
and from the definition of the surface response function in eq.~\eqref{def_g_ph}, we recover eq.~\eqref{gspecular}. 

\subsection{Water-solid interface}

So far, we have discussed the surface response of a solid exposed to vacuum. However, in order to evaluate effective Coulomb interactions in nanochannels according to eq.~\eqref{main_result}, we require the response of the solid in contact with a water slab. The generalization is straightforward if the water is described as a local dielectric medium with permittivity $\epsilon_w$. Then, if the solid is also a local dielectric (see Sec. III.A), the boundary conditions in eq.~\eqref{BC_local} become 
\begin{equation}
\begin{split}
&\phi_{\rm ext} + \phi_{\rm ind} = \phi_m \\
&\epsilon_w(\phi_{\rm ext} - \phi_{\rm ind}) = \epsilon_m \phi_m,
\end{split}
\label{BC_local_mod}
\end{equation}
so that
\beq
g(q) = \frac{\epsilon_m - \epsilon_w}{\epsilon_m + \epsilon_w} .
\label{glocal}
\eeq

For an arbitrary solid material in the SR approximation, the boundary conditions in eq.~\eqref{BC_specular} are modified in a similar way, and one obtains 
\beq
g (q,\omega) = \frac{1-\epsilon_w q\ell_q(\omega)}{1+ \epsilon_w q\ell_q(\omega)}.
\label{gSR_water}
\eeq

We may further extend these results to the case where water has anisotropic permittivity, as is typically the case in nanoscale confinement~\cite{Schlaich2016,Fumagalli2018,Bonthuis2011}. In the absence of polarization within the solid wall, the potential created by a point charge $+e$ in anisotropic water with a permittivity tensor $\overline{\overline \epsilon}$ satisfies the Poisson equation 
\begin{equation}
\nabla \left( \overline{\overline \epsilon} \cdot \nabla \phi \right) = -\frac{e}{\epsilon_0} \delta (\rr ). 
\end{equation}
For a charge placed at $(\rho = 0, z = 0)$, this is solved by 
\begin{equation}
\phi_{\rm ext} (\rho,z) = \frac{e}{4 \pi \epsilon_0\sqrt{\epsilon_{\parallel} \epsilon_{\perp}} \sqrt{\rho^2 + \frac{\epsilon_{\parallel}}{\epsilon_{\perp}} z^2}},
\label{phiext_aniso} 
\end{equation}
where $\epsilon_{\perp}$ (resp. $\epsilon_{\parallel}$) is the component of the permittivity tensor in the confined (resp. non-confined) direction. Eq.~\eqref{phiext_aniso} becomes, after Fourier transformation, 
\begin{equation}
\phi_{\rm ext} (q,z) = \frac{e}{4\pi \epsilon_0 \sqrt{\epsilon_{\parallel} \epsilon_{\perp}} } \frac{2\pi}{q} e^{-aq|z|},
\end{equation}
with $a = \sqrt{\epsilon_{\parallel}/\epsilon_{\perp}}$: this now replaces eq.~\eqref{phiext} for the external potential applied on the confining walls. Hence, taking into account the dielectric anisotropy amounts to replacing $\epsilon_w \mapsto  \sqrt{\epsilon_{\parallel} \epsilon_{\perp}}$, and introducing factors $a$ in all the exponentials of the type $e^{-qz}$. In particular, eq.~\eqref{main_result} for the total potential in the channel midplane becomes
\begin{equation}
\phi_{\rm tot}(q,0) =  \frac{e}{4\pi \epsilon_0\sqrt{\epsilon_{\parallel} \epsilon_{\perp}} } \frac{2\pi}{q} \left( 1- \frac{2 g_m(q) e^{-aqh}}{1+g_m(q)e^{-aqh}} \right). 
\label{phitotaniso}
\end{equation}
The surface response function of the dielectric solid is modified according to 
\begin{equation}
g(q) = \frac{\epsilon_m -  \sqrt{\epsilon_{\parallel} \epsilon_{\perp}} }{\epsilon_m +  \sqrt{\epsilon_{\parallel} \epsilon_{\perp}} },
\label{ganiso}
\end{equation}
and in the SR approximation for an arbitrary solid 
\beq
g (q,\omega) = \frac{1-\sqrt{\epsilon_{\parallel} \epsilon_{\perp}} q\ell_q(\omega)}{1+ \sqrt{\epsilon_{\parallel} \epsilon_{\perp}} q\ell_q(\omega)}.
\eeq

\section{Confined potential for model wall materials}

In this section we use the results of sections II and III to discuss explicitly the nature of the confined Coulomb interactions for different channel wall materials. 

\subsection{Local dielectric: quasi-2D Coulomb interactions}

A situation typically encountered in nanofluidics is the one of a channel with insulating walls that can be described by a local dielectric constant, that is much lower than that of water. Since it applies to biological nanopores, such a configuration has been extensively studied in a 1D geometry~\cite{Teber2005,Levin2006,Loche2019,Kavokine2019}, and the results have recently been extended to a 2D geometry~\cite{Robin2021} through a direct solution of Poisson's equation. The formal result obtained in ref.~\cite{Robin2021} can be recovered without computation in the surface response function framework. By simply replacing eq.~\eqref{glocal} into eq.~\eqref{main_result}, we obtain the Fourier-transformed Coulomb potential as 
\beq
\phi_{\rm tot} (q) = \frac{e}{2 \epsilon_0 \epsilon_w q} \left( 1- \frac{2 (\epsilon_m - \epsilon_w ) e^{-qh}}{\epsilon_m+\epsilon_w + (\epsilon_m - \epsilon_w) e^{-qh}} \right).
\label{hankquasi2D}
\eeq
The real-space potential can then be obtained by \PR{performing the inverse Fourier transform} according to eq.~\eqref{invhankel}. \PR{Expanding the last term of eq.~\eqref{hankquasi2D} in a geometric series, we obtain
\beq
\phi_{\rm tot}(\rho) = \frac{e}{4 \pi \epsilon_0\epsilon_w} \left(\frac 1 \rho + 2\sum_{n=1}^{+ \infty} \left( \frac{\epsilon_w - \epsilon_m}{\epsilon_w + \epsilon_m}\right)^n\frac{1}{\sqrt{\rho^2 + h^2 n^2}} \right).
\label{realspace_quasi2D}
\eeq
The infinite sum in the above equation can be interpreted in terms of the induced electrostatic potential $\phi_{\rm ind}(\rho)$, created by all the image charges.} The result of eq.~\eqref{realspace_quasi2D} is plotted in Fig. 2a for a channel height $h = 2~\rm nm$. For simplicity, we assume that the channel is filled with water that has an isotropic dielectric constant $\epsilon_w$, and the wall material has a permittivity $\epsilon_m = 2$. Similarly to the case of a 1D nanotube~\cite{Kavokine2021}, the behavior of the potential as a function of the distance $\rho$ from the charge can be split into three regions. At short distances $\rho \ll h$, the test charge only "sees" the dielectric response of water, and $\phi(\rho) \sim 1/\epsilon_w \rho$. Conversely, at large distances $\rho \gg h$, the potential is mostly screened by the walls, and $\phi(\rho) \sim 1/\epsilon_m\rho$. At intermediate distances, there is a regime where the electric field lines remain parallel to the channel walls due to the dielectric contrast $\epsilon_w \gg \epsilon_m$, and the potential has a logarithmic behavior as a function of distance. These three limiting regimes can be captured in the following analytical expression~\cite{Robin2021}: 
\begin{equation}
\phi(\rho) = \frac{e \mathcal{K}}{2\pi \epsilon_0 \epsilon_w h} \log \left( \frac{\rho + \xi}{\rho} \right), 
\label{log_potential}
\end{equation}
where $\mathcal{K} \approx 1.1$ is a geometrical factor, and $\xi = \epsilon_w h /(2\epsilon_m)$ is a lengthscale that sets the transition between the intermediate distance logarithmic and the long-distance $1/\rho$ regimes. This expression is plotted in Fig.~2a and is in excellent agreement with the exact solution.

\begin{figure*}
\centering
\includegraphics[width=0.8\textwidth]{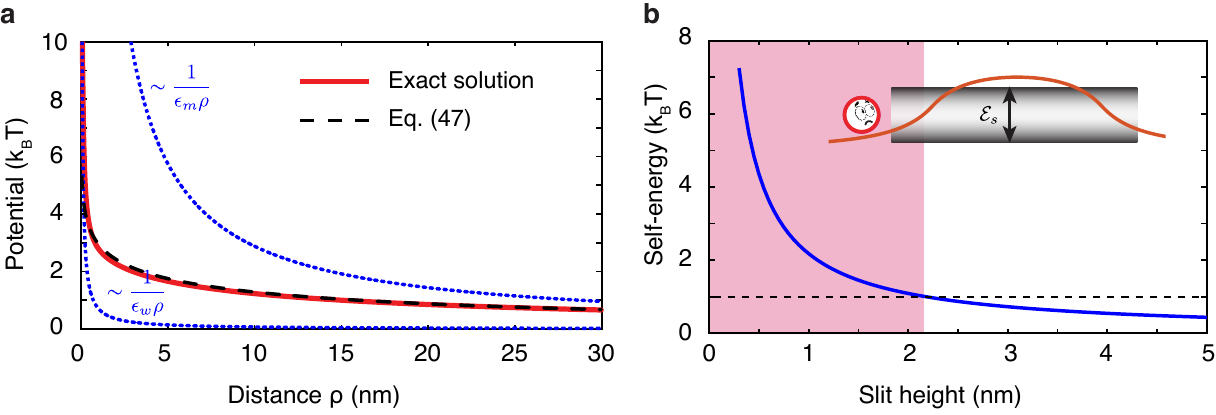}
\caption{Confined potential in a 2D channel ($h = 2~\rm nm$) with insulating walls ($\epsilon_m = 2$). \textbf{a.} Potential along the channel axis, compared to its limiting expressions at short and long distances. \textbf{b.} Self-energy barrier as a function of channel height. The shaded region corresponds to channel radii for which the self-energy barrier is greater than $k_{\rm B} T$.}
\label{interactions_2Dion}
\end{figure*}

The physical effects of interaction confinement with dielectric walls have been discussed, in the 1D channel case, in ref.~\cite{Kavokine2021}. We obtain a similar picture for a 2D channel, the essential point being that the effective potential in Fig. 2a is always larger than the bulk water Coulomb potential $\phi(\rho) = 1/4\pi \epsilon_0 \epsilon_w \rho$: confinement leads to enhanced Coulomb interactions. The significance of this interaction enhancement can be assessed by computing the corresponding self-energy (discussed in the Introduction): it can be obtained as $\mathcal{E}_s = e\phi_{\rm ind}(0,0)/2$. $\mathcal{E}_s$ is plotted as a function of channel height in Fig. 2b. It is found to be greater than $k_{\rm B}T$ for $h\lesssim 2~\rm nm$, which roughly establishes a threshold for the importance of interaction confinement effects in 2D geometry.

\subsection{Perfect metal}

We now consider the opposite limit for the dielectric behavior of the channel wall material: a perfect metal, defined by $\epsilon_m \to \infty$. Such a model has been applied to nanopores in carbon electrodes~\cite{Merlet2012,Lee2014}, and several computations of the corresponding effective Coulomb interactions, based on the solution of Poisson's equation, have been proposed both in 1D and 2D geometry~\cite{Weber1939,Kondrat2011,Loche2019}. In the framework of surface response functions, $\epsilon_m \to \infty$ implies $g(q) = 1$, so that eq.~\eqref{main_result} becomes
\begin{equation}
\phi_{\rm tot}(q,0) =  \frac{e}{4\pi \epsilon_0\epsilon_w } \frac{2\pi}{q} \mathrm{tanh} (q h/2),
\end{equation}
and the real space potential, according to eq.~\eqref{invhankel}, is given by
\begin{equation}
 \phi(\rho) = \frac{e}{4\pi \epsilon_0\epsilon_w } \int_0^{+\infty} \d q \, J_0(q \rho) \mathrm{tanh}(qh/2). 
 \end{equation}
We may compute this integral as a series expansion. First, we introduce the notation 
 \begin{equation}
 \phi(\rho) =  \frac{e}{4\pi \epsilon_0\epsilon_w } \frac{2}{h} \, \mathcal{I}(\tilde \rho),
 \end{equation}
with $\tilde \rho \equiv 2 \rho / h$, and 
\begin{equation}
\mathcal{I}(\tilde \rho) = \int_0^{+\infty} \d q \, J_0(q\tilde \rho) \mathrm{tanh}(q). 
\label{ImetalLaplace}
\end{equation}
Then, we make use of the property 
\begin{equation}
\mathcal{I}(\tilde \rho) = \int_0^{+\infty} \d s \, \mathcal{L} [J_0(q\tilde \rho)](s) \mathcal{L}^{-1}[\mathrm{tanh}(q)](s), 
\end{equation}
where $\mathcal{L}$ is the Laplace transform. For the Bessel function, we have $\mathcal{L} [J_0(q\tilde \rho)](s) = 1/\sqrt{\tilde \rho^2 + s^2}$. The hyperbolic tangent has poles at $iq_n = i(2n+1)\pi/2, n \in \mathbb{Z}$ on the imaginary axis. Hence, its inverse Laplace transform is given by 
\begin{equation}
\mathcal{L}^{-1}[\mathrm{tanh}(q)](s) = \frac{1}{2i\pi} \int_{\delta - i \infty}^{\delta + i \infty} \d q \, \tanh(q) e^{qs},
\end{equation}
with $\delta >0$. This integral is computed by closing the integration path in the left complex plane, and making use of the Cauchy residue theorem: 
\begin{equation}
\mathcal{L}^{-1}[\mathrm{tanh}(q)](s) = \sum_{iq_n}  \mathrm{Res}_{q= iq_n}[\tanh(q) e^{qs}],
\end{equation}
Since the residue of the hyperbolic tangent at the poles $iq_n$ is 1, we obtain 
\begin{align}
\mathcal{L}^{-1}[\mathrm{tanh}(q)](s) &= \sum_{n=-\infty}^{+\infty}  e^{i(2n+1)\pi s/2} \\ &= 2 \sum_{n=0}^{+\infty} \cos \left( \frac{2n+1}{2} \pi s \right).
\end{align}
Replacing into eq.~\eqref{ImetalLaplace} yields
\begin{equation}
\mathcal{I}(\tilde \rho) = 2 \sum_{n=0}^{+\infty} \int_0^{+\infty} \d s \frac{\cos ((2n+1)\pi s/2)}{\sqrt{\tilde \rho^2+s^2}}. 
\end{equation}
Here, we may recognize the integral representation $K_0$, the modified Bessel function of the second kind of order 0: 
\begin{equation}
\mathcal{I}(\tilde \rho) = 2 \sum_{n=0}^{+\infty} K_0 \left( \frac{2n+1}{2} \pi \tilde \rho \right). 
\end{equation}
\normalsize
Finally, we obtain for the potential in the perfect metal limit 
\begin{equation}
 \phi(\rho) = \frac{e}{\pi \epsilon_0\epsilon_w h } \sum_{n=0}^{+\infty} K_0 \left( (2n+1) \pi  \frac{\rho}{h} \right). 
\end{equation}
This result differs from the one given 
by Kondrat and Kornyshev~\cite{Kondrat2011}, which does not reduce to an unperturbed Coulomb potential in the limit $\rho \to 0$. 
However, in the (interesting) limit $\rho \gg h$, we recover the same asymptotic form as in ref.~\cite{Kondrat2011}: 
\begin{equation}
\phi(\rho) \approx \frac{e}{\pi \epsilon_0\epsilon_w} \frac{e^{-\pi \rho/h}}{\sqrt{2 h \rho}}.
\label{metal_scaling}
\end{equation}
This limiting expression reveals that the metallic walls produce exponential screening of the confined Coulomb potential. We leave a complete discussion of the confined potential until after we introduce the Thomas-Fermi model for the channel walls, which interpolates between a metallic and an insulating behavior. 

\subsection{Thomas-Fermi model}

\begin{figure*}
\centering
\includegraphics[width=0.8\textwidth]{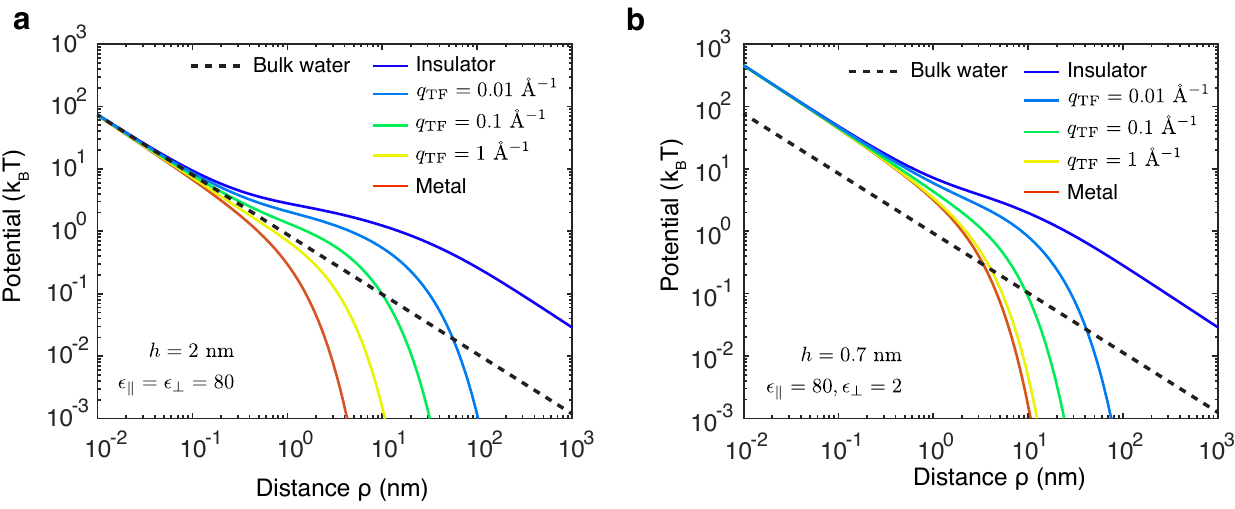}
\caption{Coulomb potential created by an ion of charge $e$ inside a nano-slit, for different models of the confining walls' dielectric response. \textbf{a}. Slit of height $h = 2~\rm nm$ and water with isotropic permittivity ($\epsilon_w = 80$). \textbf{b}. Slit of height $0.7~\rm nm$ and water with anisotropic dielectric response ($\epsilon_{\parallel} = 80, \epsilon_{\perp} = 2$). The background dielectric constant is set to $\epsilon_m = 2$ in all instances.}
\label{ionic_interactions_TF}
\end{figure*}

In order to continuously explore the range of screening properties between metal and insulator, we make use of the Thomas-Fermi model, which introduces a non-local form of the bulk dielectric function for the channel wall material~\cite{Mahan}: 
\begin{equation}
\epsilon(q,q_z) = \epsilon_m + \frac{q_{\rm TF}^2}{q^2+q_z^2}.
\label{epsilon_TF}
\end{equation}
Here $\epsilon_m$ is the background dielectric constant that accounts for screening by high-energy optical transitions, and $q_{\rm TF}$ is the Thomas-Fermi wavevector. Qualitatively, the Thomas-Fermi model introduces a screening length for the potential: $\lambda_{\rm TF} = q_{\rm TF}^{-1}$. In an insulator, $q_{\rm TF} = 0$ and the screening length is infinite, while in a perfect metal $q_{\rm TF} \to \infty$ and the potential is screened over an infinitely short distance below the surface. The reported computations\cite{Vorotyntsev1980,Kornyshev1980,Kornyshev1982,Kaiser2017} of Coulomb interactions above a single Thomas-Fermi surface -- that rely, again, on directly solving Poisson's equation -- are quite analytically involved, and, to our knowledge, no evaluation of the Coulomb potential in between two Thomas-Fermi walls has been proposed. 

 The surface response function framework is particularly powerful for this purpose. Indeed, applying the SR approximation (eq.~\eqref{gSR_water}) to the dielectric function in eq.~\eqref{epsilon_TF} yields the surface response function in the Thomas-Fermi model: 
\begin{equation}
g(q) = \frac{\epsilon_m f_{\rm TF}(q) -  \epsilon_w}{\epsilon_m f_{\rm TF}(q) +  \epsilon_w }, ~~~ f_{\rm TF}(q) = \sqrt{1+\frac{1}{\epsilon_m} \frac{q_{\rm TF}^2}{q^2}}. 
\end{equation}
Replacing this into eq.~\eqref{main_result} directly yields the confined Coulomb potential in Fourier space. No convenient analytical expression can be given in this case for the potential in real space, which is obtained in practice by numerical integration following eq.~\eqref{invhankel}. \LB{This potential can however be implemented in molecular simulations using an appropriate fit of the numerically estimated potential, see next section and eq.(\ref{eqn:yukamod}). }

In Fig. 3, we plot the Coulomb potential created by an ion in a slit-like channel for different values of the Thomas-Fermi wavevector $q_{\rm TF}$ of the confining walls. We consider two different cases: a channel of height $h = 2~\rm nm$ filled with isotropic water (permittivity $\epsilon_w = 80$), and a channel of height $h = 0.7 ~\rm nm$ filled with anisotropic water ($\epsilon_{\parallel} = 80, \epsilon_{\perp} = 2$)~\cite{Schlaich2016}. In both cases, we assume a background dielectric constant $\epsilon_m = 2$. Note that in the anisotropic case, we generalize our results as explained in Sec. III D. As soon as the wall material has conduction electrons (that is, $q_{\rm TF}$ is non-zero), the potential becomes exponentially screened at long distances. If the channel height $h \gg \lambda_{\rm TF}$, the finite size of the screening cloud within the wall plays no significant role and the perfect metal limit applies: at distances $\rho \gtrsim h$, the potential is exponentially screened over the lengthscale  $(h/\pi) \sqrt{\epsilon_{\parallel}/\epsilon_{\perp}}$, as given by eq.~\eqref{metal_scaling}. But if $h \lesssim \lambda_{\rm TF}$, the potential is screened over a longer lenghtscale of order $\lambda_{\rm TF}$: as soon as the screening length is comparable to the channel size, the perfect metal model underestimates the confined Coulomb potential. At distances that are too short for the exponential screening to apply, the material behaves as an insulator with the background dielectric constant $\epsilon_m$. At large enough distances, the confined potential always becomes smaller than the bulk Coulomb potential. However, at small distances, Coulomb interactions may still be enhanced with respect to the bulk, depending on the screening length and dielectric anisotropy of water. This suggests a rich dependence of confined ion transport on the channel walls' electronic properties. 

\LB{As a last remark, } \PR{in the above discussion, we only considered the electrostatic potential within the midplane $z=0$ of the nanochannel. However, the potential can be evaluated numerically (and analytically in the dielectric case) for any value of $z$, and is found to vary by 10 \% at most across a channel of height $h = 7~\angstrom$.}

\section{Tuning Wien effect ion transport with interaction confinement}

In this section, \LB{we illustrate the principle of interaction confinement by studying the transport of ions confined in 2D slits}. We demonstrate using molecular dynamics (MD) simulations that the electronic properties of a channel wall can impact the ion transport within the channel. Typically, the electronic properties of a solid can only be accounted for in computationally expensive ab initio MD; in the framework of classical MD, there are no electronic degrees of freedom (although methods for incorporating Thomas-Fermi screening within classical MD have been proposed~\cite{Scalfi2020,Schlaich2022}). Here, we make use instead of the much less expensive implicit solvent brownian dynamics simulations (Fig. 4a), where the effective ion-ion interactions computed in the previous section can be directly implemented. 

\begin{figure}
	\centering
	\includegraphics[width=0.45\textwidth]{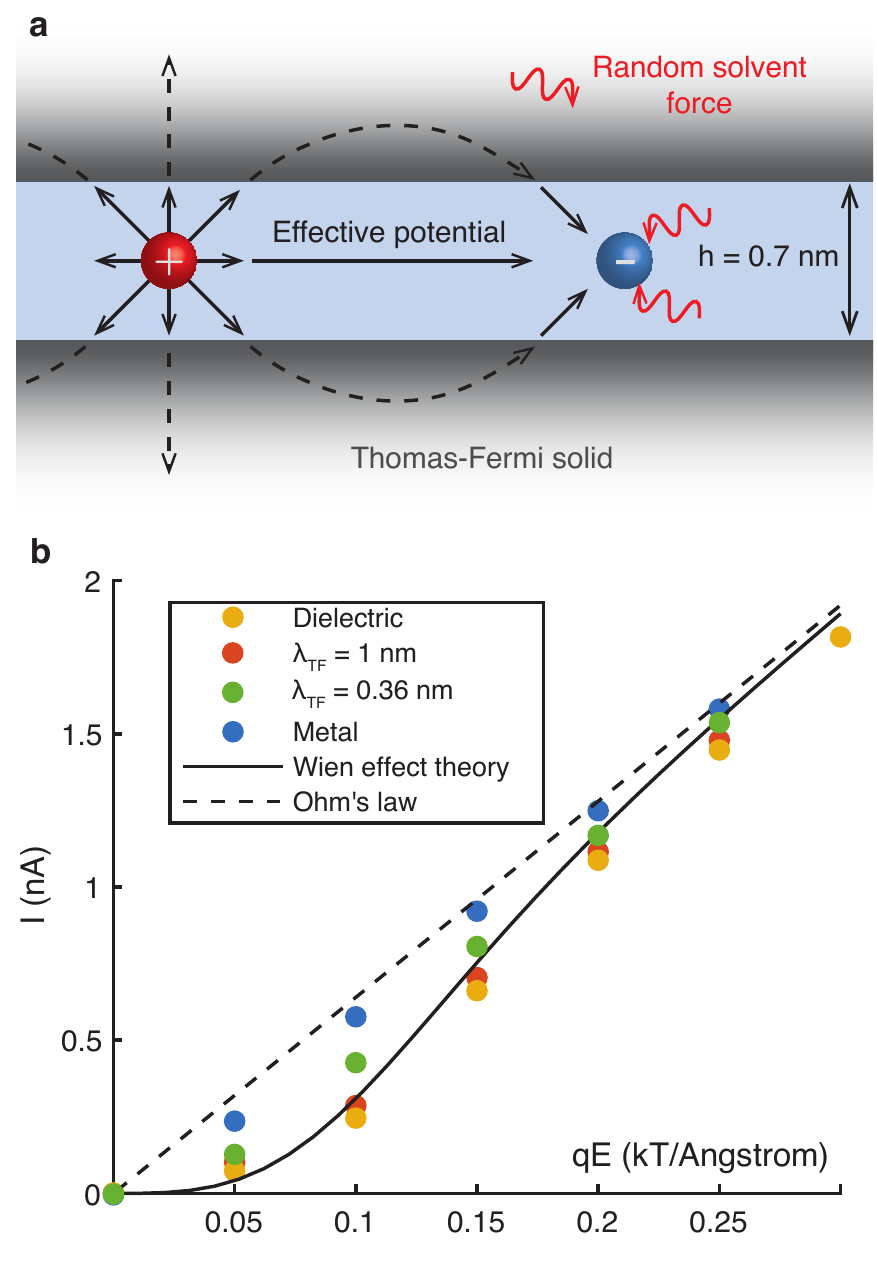}
	\caption{\textbf{a}. Illustration of the brownian dynamics simulation setup. Point-like ions restricted to move in two dimensions interact through an effective potential and are subject to a random force from the implicit solvent. \textbf{b}. Current-voltage characteristics as obtained from the brownian dynamics simulations. The characteristics display a Wien effect non-linearity\PR{, except in the case of metallic walls, where the linear Ohm's law is recovered. The solid line corresponds to the analytical prediction for ionic conduction through a dielectric nanochannel, according to ref. \cite{Robin2021}.} }
\end{figure}

\subsection{Simulation methods}

We use implicit solvent overdamped Langevin dynamics implemented in the open-source LAMMPS software~\cite{Plimpton1995}. Our simulation box typically contains 100 ions of each sign from a divalent salt  with diffusion coefficient $D = 10^{-9} \, \si{m^2 \cdot s^{-1}}$, confined in a nanochannel with height $h = 0.7 \, \si{nm}$ and lateral dimensions $200 \, \si{nm} \times 200 \, \si{nm}$. Ions are restricted to move within the center plane $z=0$ of the channel, with periodic boundary conditions in both $x$ and $y$. We apply a constant electric field in the $x$ direction and measure the resulting ionic current, under a Langevin thermostat at $T = 298 \, \si{K}$. Our simulation procedure is summarized in Fig. 4a. The simulation is let to run for $1 \, \si{ns}$ of physical time. 

The ions interact with the effective confined interactions corresponding to channel walls described within the Thomas-Fermi model. Since there is no convenient analytical form for the confined potential in real space, we fit the result of the numerical integration in eq.~\eqref{invhankel} with the following analytical form 
\PR{\begin{equation}
	\mathcal U(\rho) =  \frac{A}{\rho} \left( 1 - \alpha e^{-\kappa \rho}\right) e^{- q_e \rho}, 
	\label{eqn:yukamod}
\end{equation}
where  $A, \kappa, \alpha$ and $q_e$ are adjustable parameters. In the case of insulating walls, there is no exponential screening of the potential at long distance, so that $q_e = 0$ and the above expression can be thought of as an interpolation between the two limiting behaviors of the potential: 
\begin{equation}
	\mathcal{U} \sim  \begin{cases}
		\frac{e^2}{4 \pi \epsilon_0 \sqrt{\epsilon_\parallel \epsilon_\bot} \rho} \quad \text{for} \quad \rho \to 0\,\\ 
		\frac{e^2}{4 \pi \epsilon_0 \epsilon_m \rho} \quad \text{for} \quad \rho \to + \infty.
	\end{cases}
\end{equation}
In practice, $A,\kappa$ and $\alpha$ are fitted so that the transition between the two regimes reproduces the exact numerical result for the potential. Ultimately, for a material with finite screening length, $q_e$ is fitted to reproduce the exponential decay of the potential at long distance. In the simulations, we also \LB{add} a \LB{short-distance} term, $A(1-\alpha) e^{-r/r_0}/r$, into the potential's \LB{expression} to avoid the divergence at short distance (with $r_0 = 1~\textrm{nm}$, of the order of the minimal approach distance between two ions). 

Numerically, we use $\epsilon_{\parallel} = 80$ and $\epsilon_{\perp} = 2$ for the components of the water dielectric permittivity and $\epsilon_m = 2$ for the wall's background dielectric constant. The corresponding values of the parameters are given in Table I.  }

\begin{table}
	\centering
	\begin{tabular}{ccccc}
		Material & $A~(\si{kcal \cdot \angstrom \per \mole})$ & $\alpha$ & $\kappa~(\si{\nm^{-1}})$  & $q_e ~(\si{\nm^{-1}})$\\
		\hline
	
		Dielectric & $134$ & $0.81$ & $0.16$& $0$\\
		$q_\text{TF} = 1~\si{\nm^{-1}}$ & $88$ &  $0.72$& $0.14$& $0.18$\\
		$q_\text{TF} = 2.8~\si{\nm^{-1}}$ & $141$ &  $0.80$& $0.05$& $0.28$ \\
		Metal & $28.7$ & $0$ & --&$0.46$\\
		\hline
	\end{tabular}
	\caption{Values of the parameters in eq.~\eqref{eqn:yukamod} used for our simulations.}
\end{table}

\subsection{Results}

We present in Fig. 4b the current-voltage characteristics of 2D channels ($h = 7~\rm \angstrom$), as obtained from our brownian dynamics simulations, \PR{for four different wall materials characterized by increasing values of the Thomas-Fermi screening length: $\lambda_{\rm TF} = 0$ (perfect metal), $\lambda_{\rm TF} = 0.36 ~\rm nm$, $\lambda_{\rm TF} = 1~\rm nm$ and $\lambda_{\rm TF} = \infty$ (insulator). We find that, in all but the perfect metal cases,} the measured current  displays a non-linear dependence on the applied electric field and  the channel conductance is \LB{hindered} with respect to the Ohm's law prediction: this is a signature of the second Wien effect. The Wien effect has been historically known to govern the conduction in weak electrolytes; it has recently been predicted to occur as well in strong electrolytes \LB{when confined} in nanoscale channels, because oppositely charged ions \LB{form Bjerrum pairs} due to the reinforced Coulomb interactions in confinement~\cite{Kavokine2019,Robin2021}. In refs.~\cite{Kavokine2019,Robin2021}, the Wien effect was studied in the case of insulating channel walls. Here, we find that \LB{the Wien effect can persist even in the presence of electronic screening and} the magnitude of the  effect can be tuned by the channel wall's electronic properties. 
\LB{As shown in Fig.4b, the voltage-current characteristics is found to depart from the linear Ohm's law at small applied electric field and the effect is largest when
the TF screening length is larger than the channel height.}
\LB{These results are further compared  with the theoretical predictions from ref. \cite{Robin2021}, which are based on the modelling of the Bjerrum pair dynamics in 2D under the application of an electric field. The analytical result -- Eq.(5) in ref. \cite{Robin2021} -- is shown as a solid line in Fig.4b -- we stress that this prediction includes no fitting parameter as it only incorporates the value of the dielectric constant, channel height and ion concentration --.} 
The conductance is found to be in very good agreement with the theoretical prediction, when the TF screening length is larger than the channel height. But the non-linearity is weakened as soon as the screening length becomes comparable to the channel height: the screening then modifies the potential at short enough distances to affect the binding energy of the pairs. In the perfect metal case, the screening essentially destroys the Bjerrum pairs and the current collapses onto the Ohm's law prediction.

\section{Discussion and conclusions}

In this paper, we have introduced the broad notion of \emph{interaction confinement}, defined as the regime where the interactions between particles inside a channel are affected by the nature of the channel wall. \PR{Focusing} on a two-dimensional channel geometry, we developed a new theoretical framework that allowed us to compute these modified interactions, typically for the case of ions within a confined electrolyte. Our framework is based on \emph{surface response functions}, that describe a solid surface's response to an external potential. It is hence very general: the confined interactions can be computed given any channel wall material, provided that its surface response function is known. We evaluate in particular the confined interactions for channel walls described within the Thomas-Fermi model, for which no expression was available in the literature. 

\PR{While our approach is limited to the 2D geometry, it allows us to shed some light on the properties of confined ions in general. We expect that most qualitative results to extend to other geometries, such as 1D tube-like channels. In other words, 2D nanochannels represent a model platform to explore the consequences of interaction confinement in ion and water transport.}

Our framework can be used to estimate Coulomb interactions in experimentally accessible nanofluidic channels. For instance, graphite can be described as a Thomas-Fermi conductor with $\lambda_{\rm TF} = 1.2~\rm nm$ (ref.\cite{Miyazaki2008}) and $\epsilon_m \sim 4$ (ref.\cite{Hwang2007}), and boron nitride as an insulator with $\epsilon_m = 6$ (ref.\cite{Geick1966}). These estimates are important for predicting ion transport properties within such channels: we found indeed (Sec. V) that even simple observables such as current-voltage characteristics are affected by the channel wall's dielectric screening properties, that need to be treated beyond a simple local approximation. 

The limitation of our present results lies mainly in the local dielectric treatment for water. Although we do take into account the anisotropy induced by confinement, a more rigorous treatment involving the non-local dielectric response will be the subject of future work, as it may introduce corrections for the narrowest channels. Furthermore, our theory uses as an input the channel wall's surface response function \emph{in the presence of water}. If this response function is to be computed from first principles, renormalization of the wall's electronic properties by the presence of water should be taken into account, for example in the spirit of refs.~\cite{Misra2021,Robert2022}. 

Ultimately, we would like to emphasize that the notion of interaction confinement is not restricted to confined ion transport. For instance, the dynamics of confined water are determined by the Coulomb interactions between the water molecules. These interactions are also subject to screening by the nearby solid wall and depend even more subtly on its electronic properties: indeed, molecular scale water fluctuations are faster than ionic motion, and these may couple not only to static, but also to dynamical screening properties. In this way, the fluctuation-induced quantum friction phenomenon~\cite{Kavokine2022} can be seen as an effect of interaction confinement on water: it results essentially from the Coulomb interactions in water being dynamically screened by the solid's electronic excitations. Altogether, interaction confinement appears as a fundamental feature of fluid transport at the nanoscale.

\begin{acknowledgments}
The Flatiron Institute is a division of the Simons Foundation.
L.B. acknowledges funding from the EU H2020 Framework Programme/ERC Advanced Grant agreement number 785911-Shadoks. 
This work was granted access to the HPC resources of CINES under the allocation A0090710395 made by GENCI.
\end{acknowledgments}

\section*{Data Availability Statement}

The data that support the findings of this study are available from the corresponding author upon reasonable request.

\end{document}